\documentclass[aps,twocolumn,prl,superscriptaddress,longbibliography,reprint]{revtex4-2}
\usepackage{amsfonts}
\usepackage{mathrsfs}
\usepackage{amsmath}
\usepackage{color}
\usepackage{graphicx}
\usepackage{bm}
\usepackage{amssymb}
\usepackage{xspace}
\usepackage{epstopdf}
\usepackage{longtable}
\usepackage{multirow}
\usepackage{booktabs}
\usepackage{textcomp}
\usepackage[title]{appendix}
\usepackage{float}
\usepackage{dsfont}
\usepackage[colorlinks=true, letterpaper=true, pdfstartview=FitV, linkcolor=blue, citecolor=blue, urlcolor=blue]{hyperref}
\usepackage[]{changes}

\providecommand{\U}[1]{\protect\rule{.1in}{.1in}}

\begin{document}

\title{Orbital Origin of Intrinsic Planar Hall Effect}

\author{Hui Wang}
\thanks{These authors contributed equally to this work.}
\affiliation{Research Laboratory for Quantum Materials, Singapore University of Technology and Design, Singapore 487372, Singapore}
\affiliation{Division of Physics and Applied Physics, School of Physical and Mathematical Sciences, Nanyang Technological University, Singapore 637371, Singapore}

\author{Yue-Xin Huang}
\thanks{These authors contributed equally to this work.}
\affiliation{Research Laboratory for Quantum Materials, Singapore University of Technology and Design, Singapore 487372, Singapore}
\affiliation{School of Sciences, Great Bay University, Dongguan 523000, China}
\affiliation{Great Bay Institute for Advanced Study, Dongguan 523000, China}

\author{Huiying Liu}
\affiliation{School of Physics, Beihang University, Beijing 100191, China}
\affiliation{Research Laboratory for Quantum Materials, Singapore University of Technology and Design, Singapore 487372, Singapore}

\author{Xiaolong Feng}
\affiliation{Research Laboratory for Quantum Materials, Singapore University of Technology and Design, Singapore 487372, Singapore}
\affiliation{Max Planck Institute for Chemical Physics of Solids, N\"{o}thnitzer Strasse 40, D-01187 Dresden, Germany}

\author{Jiaojiao Zhu}
\affiliation{Research Laboratory for Quantum Materials, Singapore University of Technology and Design, Singapore 487372, Singapore}

\author{Weikang Wu}
\affiliation{Key Laboratory for Liquid-Solid Structural Evolution and Processing of Materials, Ministry of Education, Shandong University, Jinan 250061, China}
\affiliation{Research Laboratory for Quantum Materials, Singapore University of Technology and Design, Singapore 487372, Singapore}

\author{Cong Xiao}
\email{congxiao@um.edu.mo}
\affiliation{Institute of Applied Physics and Materials Engineering, University of Macau, Macau, China}
\affiliation{Department of Physics, The University of Hong Kong, Hong Kong, China}
\affiliation{HKU-UCAS Joint Institute of Theoretical and Computational Physics at Hong Kong, Hong Kong, China}

\author{Shengyuan A. Yang}
\affiliation{Institute of Applied Physics and Materials Engineering, University of Macau, Macau, China}

\begin{abstract}
Recent experiments reported an antisymmetric planar Hall effect, where the Hall current is odd in the in-plane magnetic field and scales linearly with both electric and magnetic fields applied. Existing theories rely exclusively on a spin origin, which requires spin-orbit coupling to take effect. Here, we develop a general theory for the intrinsic planar Hall effect (IPHE), highlighting a previously unknown orbital mechanism and connecting it to a band geometric quantity --- the anomalous orbital polarizability (AOP). Importantly, the orbital mechanism does not request spin-orbit coupling, so sizable IPHE can occur and is dominated by orbital contribution in systems with weak spin-orbit coupling. Combined with first-principles calculations, we demonstrate our theory with quantitative evaluation for
bulk materials $\mathrm{TaSb_{2}}$, $\mathrm{NbAs_{2}}$, and $\mathrm{SrAs_{3}}$. We further show that AOP and its associated orbital IPHE can be greatly enhanced at topological band crossings, offering a new way to probe topological materials.
\end{abstract}
\maketitle


The Hall effects are of fundamental importance in condensed matter physics~\cite{QHE1986,Nagaosa2010,Sinova2015}. In nonmagnetic materials, Hall effect appears under an applied magnetic field. For $B$ field out of the transport plane, i.e., the plane formed by the driving $E$ field and the measured Hall current $j_\text{H}$, this is the ordinary Hall effect due to Lorentz force~\cite{Solid_1976}. Recently, a new type of Hall effect was found in experiments on certain bulk nonmagnetic materials, where a Hall response was induced by an \emph{in-plane} $B$ field and scales as $j_\text{H}\sim EB$~\cite{Ong2018,Zhou2022}. Note that this effect is distinct from many previously reported planar Hall effects~\cite{Davis1954,Tang2003,Seemann2011,Nandy2017,Ando2017,Ong2018PRX,Kumar2018,Wang2018,Ma2019,Deng2019}, where the Hall current is an even function in $B$ and hence is not a genuine Hall response but rather represents an off-diagonal anisotropic magnetoresistance \cite{Ong2018}.

There have been theoretical studies on this effect \cite{Zyuzin2020,Ortix2021,Culcer2021}. However, the current understanding is far from complete for the following reasons. First, existing theories are exclusively based on spin-orbit coupling (SOC) and Zeeman coupling between spin and in-plane $B$ field. The predicted response \emph{vanishes} when SOC is neglected. Such a treatment misses the orbital degree of freedom of Bloch electrons, which also couples to $B$ field \cite{Xiao2010} and can result in Hall transport \emph{regardless} of SOC. Second, previous works are mostly focused on specific models, such as Rashba model~\cite{Zyuzin2020}, modified Luttinger model~\cite{Culcer2021}, and honeycomb lattice model~\cite{Ortix2021}. It is urgent to develop a general theory that can be implemented in first-principles calculations for real materials. Third, from the scaling relation and symmetry constraint, one can easily see that the configuration allows an important \emph{intrinsic} contribution, i.e., an intrinsic planar Hall effect (IPHE), which is independent of scattering and manifests the inherent property of a material. To understand IPHE and make it a useful tool, one must clarify what intrinsic band geometric properties are underlying the effect.

In this work, we address the above challenges by formulating a general theory of IPHE. We show that besides Berry curvature and spin/orbital magnetic moment, there are two new band geometric quantities coming into play --- the anomalous spin polarizability (ASP) \cite{Xiao2023NLCISP} and the anomalous orbital polarizability (AOP).
Apart from spin contribution, we reveal a previously unknown orbital contribution (connected to AOP) to IPHE , which can generate a large response and dominate the effect in systems with weak SOC. We express the response tensor in terms of the intrinsic band structure of a material.  We clarify the symmetry character of the effect and obtain the form of response tensor for each of the 32 crystal classes. Combining our theory with first-principles calculations, we perform quantitative evaluation of IPHE in three concrete materials $\mathrm{TaSb_{2}}$, $\mathrm{NbAs_{2}}$, and $\mathrm{SrAs_{3}}$. We demonstrate that while spin and orbital contributions are comparable in $\mathrm{TaSb_{2}}$ (with relatively large SOC), the orbital mechanism completely dominates in $\mathrm{NbAs_{2}}$ and $\mathrm{SrAs_{3}}$ (with weaker SOC). In addition, for $\mathrm{SrAs_{3}}$, the large response can be further attributed to the enhanced AOP surrounding the gapped topological nodal loop at Fermi level, suggesting IPHE as a new way to probe topological band structures.

\textcolor{blue}{\textit{Band origin of IPHE.}}
We consider a general three-dimensional (3D) system.
The IPHE response can be derived within the extended
semiclassical theory \cite{Gao2014,Gao2015,Gao2019,Xiao2021OM}, which incorporates field corrections to the band
quantities. The resulting intrinsic Hall current is found to be $\bm j^\text{int}_\text{H}= - \int f_0(\tilde{\varepsilon})(\bm E\times \tilde{\bm{\Omega}})$ (details in the Supplemental Material \cite{supp}\nocite{Ma2015,Wang2022MChA,PAW_1994,VASP_1994,VASP_1996,GGA_1996,MPKmesh_1976,MBJ_1,MBJ_2,wannier_1,wannier_2,wannier_3}), where we take $e=\hbar=1$, $f_0$ is the Fermi distribution function, and the tilde in band energy $\tilde{\varepsilon}$ and Berry curvature $\tilde{\bm{\Omega}}$ indicates that these quantities include corrections by external fields \cite{Gao2014}. To compute the IPHE scaling as $\sim EB$, we just need to retain corrections to $\tilde{\varepsilon}$ and $\tilde{\bm{\Omega}}$ which are linear in $B$.

For Berry curvature, we have $\tilde{\bm{\Omega}}=\bm \Omega+\bm \Omega^B$, where $\bm \Omega$ is the (unperturbed) Berry curvature in the absence of external fields, and $\bm \Omega^B\sim B$ is the $B$-field-induced correction~\cite{supp}. One may write  $\boldsymbol{\Omega}^{B}=\nabla_{\boldsymbol{k}}\times\boldsymbol{\mathcal{A}}^{B}$ in terms of the $B$-field-induced Berry connection \cite{Gao2014,Xiao2021OM}:
\begin{equation}
\mathcal{A}_{b}^{B}\left(  \boldsymbol{k}\right)  =B_{a}[F_{ab}^{\text{S}%
}\left(  \boldsymbol{k}\right)  +F_{ab}^{\text{O}}\left(  \boldsymbol{k}%
\right)  ],
\end{equation}
where the subscripts $a$ and $b$ denote the Cartesian components, Einstein
summation convention is adopted, and the coefficients
$F_{ab}^{\text{S}}\boldsymbol{\ }$and $F_{ab}^{\text{O}}$ are the ASP and AOP,
respectively. For a particular band with index $n$, they are expressed as
\begin{equation}
F_{ab}^{\text{S}}\left(  \boldsymbol{k}\right)  =-2\operatorname{Re}%
\sum_{m\neq n}\frac{\mathcal{M}_{a}^{\text{S},nm}\mathcal{A}_{b}^{mn}%
}{\varepsilon_{n}-\varepsilon_{m}},\label{ASP}
\end{equation}
and
\begin{equation}
F_{ab}^{\text{O}}\left(  \boldsymbol{k}\right)  =-2\operatorname{Re}%
\sum_{m\neq n}\frac{\mathcal{M}_{a}^{\text{O},nm}\mathcal{A}_{b}^{mn}%
}{\varepsilon_{n}-\varepsilon_{m}}-\frac{1}{2}\epsilon_{acd}\partial
_{c}\mathcal{G}_{db}^{n}.\label{F}%
\end{equation}
Here, $\mathcal{A}_{b}^{nm}=\langle u_{n}|i\partial_{b}|u_{m}\rangle$ is the
unperturbed interband Berry connection, $\partial_{b}\equiv\partial_{k_{b}}$,
 $|u_{n}\rangle$ is the unperturbed cell-periodic Bloch state with
energy $\varepsilon_{n}$; $\boldsymbol{\mathcal{M}}^{\text{S},mn}=-g\mu
_{B}\boldsymbol{s}^{mn}$ and $\boldsymbol{\mathcal{\boldsymbol{M}}}%
^{\text{O},mn}=\sum_{\ell\neq n}(\boldsymbol{v}^{m\ell}+\delta^{\ell
m}\boldsymbol{v}^{n})\times\boldsymbol{\mathcal{A}}^{\ell n}/2$ are the
interband spin and orbital magnetic moments, respectively, with
$\boldsymbol{s}^{mn}$ ($\boldsymbol{v}^{m\ell}$) being the matrix elements of spin
(velocity) operator, $\mu_{B}$ is the Bohr magneton, and $g$ is the g-factor for spin;
$\mathcal{G}_{db}^{n}=\operatorname{Re}\sum_{m\neq n}\mathcal{A}_{d}%
^{nm}\mathcal{A}_{b}^{mn}$ is known as the quantum metric tensor \cite{QM1980}, and $\epsilon_{abc}$
is the Levi-Civita symbol.

Before proceeding, we have several comments on ASP and AOP. First, they are gauge-invariant quantities, as can be directly checked from (\ref{ASP}) and (\ref{F}). It follows that $\boldsymbol{\mathcal{A}}^{B}$ is also gauge invariant. Physically, it represents a positional shift of a Bloch wave packet induced by a $B$ field \cite{Gao2014}. Since $\mathcal{A}^{B}_b E_b$ then corresponds to an energy change of the wave packet, one can see that $F_{ab}^{\text{S}}E_{b}$ and $F_{ab}^{\text{O}}E_{b}$ actually give the anomalous spin and orbital magnetic moments induced by an $E$ field \cite{Xiao2021OM,Xiao2021Adiabatic,Xiao2023NLCISP}. This is why $F^\text{S}$ ($F^\text{O}$) is termed as ASP (AOP). Second, the two quantities have distinct dependence on SOC. ASP is allowed only when SOC is nonzero, which can be readily understood from its physical meaning discussed above. In contrast, AOP is not subjected to this constraint. This distinction leads to important consequences in the resulting IPHE, as will be discussed below. Third, one notices that the expression for AOP resembles ASP except for the second term in (\ref{F}) with the quantum metric tensor. Intuitively, the quantum metric tensor measures the distance between states at different wave vectors in Hilbert space \cite{QM1980,Gao2015}. Hence, its appearance in AOP can be understood, because unlike spin operator, the orbital moment operator is nonlocal in $k$-space.

As for the field-corrected band energy, we have (band index $n$ suppressed here) $\tilde{\varepsilon}=\varepsilon-\bm B\cdot(\boldsymbol{\mathcal{\boldsymbol{M}}}^{\text{S}%
}+\boldsymbol{\mathcal{\boldsymbol{M}}}^{\text{O}})$, where $\boldsymbol{\mathcal{M}}^\text{S}$ ($\boldsymbol{\mathcal{M}}^{\text{O}}$) is the intraband spin (orbital) magnetic moment \cite{Xiao2010}.

Substituting $\tilde{\bm{\Omega}}$ and  $\tilde{\varepsilon}$ into the expression for $\bm j^\text{int}_\text{H}$ and collecting terms of order $O(EB)$, we obtain the IPHE current
\begin{equation}\label{jj}
  j_{a}^{\text{int}}=\chi_{abc}^{\text{int}}E_{b}B_{c},
\end{equation}
with the response tensor
\begin{equation}
\chi_{abc}^{\mathrm{int}}=\int[d\boldsymbol{k}]f_{0}^{\prime}\Big[\Theta_{abc}^\text{S}(\boldsymbol{k})
+\Theta_{abc}^\text{O}(\boldsymbol{k})\Big],\label{eq-Kai}%
\end{equation}
where $[d\boldsymbol{k}]\equiv\sum_{n}d\boldsymbol{k}/\left(  2\pi\right)
^{3}$, and in the integrand of (\ref{eq-Kai}) we have explicitly separated spin and orbital contributions, with
\begin{equation}\label{TT}
  \Theta_{abc}^i(\boldsymbol{k})=v_a F_{cb}^i-v_b F_{ca}^i+\epsilon_{abd}\Omega_{d}\mathcal{M}%
_{c}^i
\end{equation}
and $i=\text{S},\text{O}$. Equations (\ref{jj}-\ref{TT}) give the general formula for the IPHE tensor.

We have the following observations. First, due to the $f_0'$ factor in (\ref{eq-Kai}), IPHE is a Fermi surface effect, as it should be. Second, as an intrinsic effect, $\chi_{abc}^{\mathrm{int}}$ is determined solely by the intrinsic band structure of a material. Particularly, our formula reveals its connection to the band geometric quantities ASP and AOP. The first two terms in (\ref{TT}) may be called the ASP (AOP) dipole, similar to the definition of Berry curvature dipole \cite{Fu2015}. Third, our theory reveals the orbital contribution to IPHE, which was not known before. As discussed, ASP and AOP have distinct dependence on SOC. As a result, sizable IPHE can still appear and is dominated by orbital contribution in systems with weak SOC. Furthermore, from Eqs.~(\ref{ASP}) and (\ref{F}), we see that like Berry curvatures, ASP and AOP are enhanced around small-gap regions in a band structure. Roughly speaking, ASP scales as $1/(\Delta\varepsilon)^2$ and AOP scales as $1/(\Delta\varepsilon)^3$, where $\Delta\varepsilon$ is the local gap. Therefore, AOP can be more enhanced than ASP with a decreased gap; and one can expect pronounced IPHE in topological semimetal states.

\textcolor{blue}{\textit{Symmetry property.}} From
Eqs.~(\ref{jj}-\ref{TT}), one  sees that $\chi_{abc}^{\mathrm{int}}$ is clearly
antisymmetric with respect to its first two indices. Therefore, we have $j_a^\text{int}E_a=0$, which indeed
describes a dissipationless
Hall current. We may define a corresponding IPHE charge conductivity $\sigma_{ab}^\text{int}\equiv\chi_{abc}^{\mathrm{int}}B_c$, which satisfies the requirement $\sigma_{ab}(B)=-\sigma_{ab}(-B)$ for a genuine Hall conductivity. In comparison, the previously studied planar Hall effect satisfies $\sigma_{ab}=\sigma_{ba}$ and  $\sigma_{ab}(B)=\sigma_{ab}(-B)$, so it represents a kind of anisotropic magnetoresistance.

By virtue of the antisymmetric character, $\chi_{abc}^{\mathrm{int}}$ can be reduced to a time-reversal
($\mathcal{T}$) even rank-two tensor%
\begin{equation}
\mathcal{X}_{dc}=\epsilon_{abd}\chi_{abc}^{\text{int}}/2. \label{tensor}%
\end{equation}
The IPHE current can be expressed as $\boldsymbol{j}%
^{\text{int}}=\boldsymbol{E}\times\boldsymbol{\sigma}^{\text{H}}$ with a $\mathcal{T}$-odd Hall pseudovector $\sigma_{d}^{\text{H}}=\mathcal{X}%
_{dc}B_{c}$. In a typical experimental setup for IPHE, the $B$ field is
applied within the transport plane, taken as $x$-$y$ plane. Then, the effect is
specified by only two tensor elements, $\mathcal{X}_{zx}$ and $\mathcal{X}_{zy}$.
Generally, it is convenient to choose a coordinate system that fits the crystal structure, which simplifies the form of
$\mathcal{X}$ tensor. In Table~\ref{symmetry}, we list the constraints of common point-group
operations on the two relevant tensor elements. One finds that IPHE is
forbidden by an out-of-plane rotation axis and by the horizontal mirror, but is allowed by in-plane rotation axes and vertical mirrors. The most general matrix forms of $\mathcal{X}$ in 32 crystallographic point groups are
presented in Supplemental Material \cite{supp}.

With the coordinate axes and $\mathcal{X}$ fixed, assume the in-plane $E$ ($B%
$)\ field makes an angle $\phi$ ($\varphi$) from the $x$ axis, i.e., $\boldsymbol{E}=E(\cos\phi
,\sin\phi,0)$ and $\boldsymbol{B}=B(\cos\varphi,\sin\varphi,0)$. Then, the IPHE
current will flow in the direction of $(-\sin\phi,\cos\phi,0)$,
with a magnitude given by
\begin{equation}
j^{\text{int}}=\mathcal{X}^{\text{H}}EB,
\end{equation}
where
\begin{equation}
\mathcal{X}^{\text{H}}=\mathcal{X}_{zx}\cos\varphi+\mathcal{X}_{zy}\sin
\varphi.
\end{equation}
One observes that as a linear-in-$E$ intrinsic Hall response, the magnitude $j^{\text{int}}$ does not depend on the $E$-field direction. Meanwhile, it does depend on the $B$-field direction and generally exhibits a $2\pi$ periodicity.

\begin{table}[ptb]
\caption{{}Constraints on the tensor elements of $\mathcal{X}$ from point
group symmetries. As $\mathcal{X}$ is time-reversal ($\mathcal{T}$) even,
symmetry operations $O$ and $O\mathcal{T}$ impose the same constraints. }%
\label{symmetry}%
\begin{tabular}
[b]{ccccc}\hline\hline
& \ \ $C_{n}^{z},$ $S_{4,6}^{z},$ $\sigma_{z}$\ \  & \ \ $C_{n}^{x},$
$S_{4,6}^{x},$ $\sigma_{x}$\ \  & \ \ $C_{n}^{y},$ $S_{4,6}^{y},$ $\sigma_{y}$
\  & \ $\ \ \ \mathcal{P}$ \ \ \ \\\hline
$\mathcal{X}_{zx}$ & $\times$ & $\times$ & $\checkmark$ & \ \ $\checkmark$\\
$\mathcal{X}_{zy}$\ \  & $\times$ & $\checkmark$ & $\times$ & \ \ $\checkmark
$\\\hline\hline
\end{tabular}
\end{table}

\begin{figure}[t]
\centering
\includegraphics[width=0.48\textwidth]{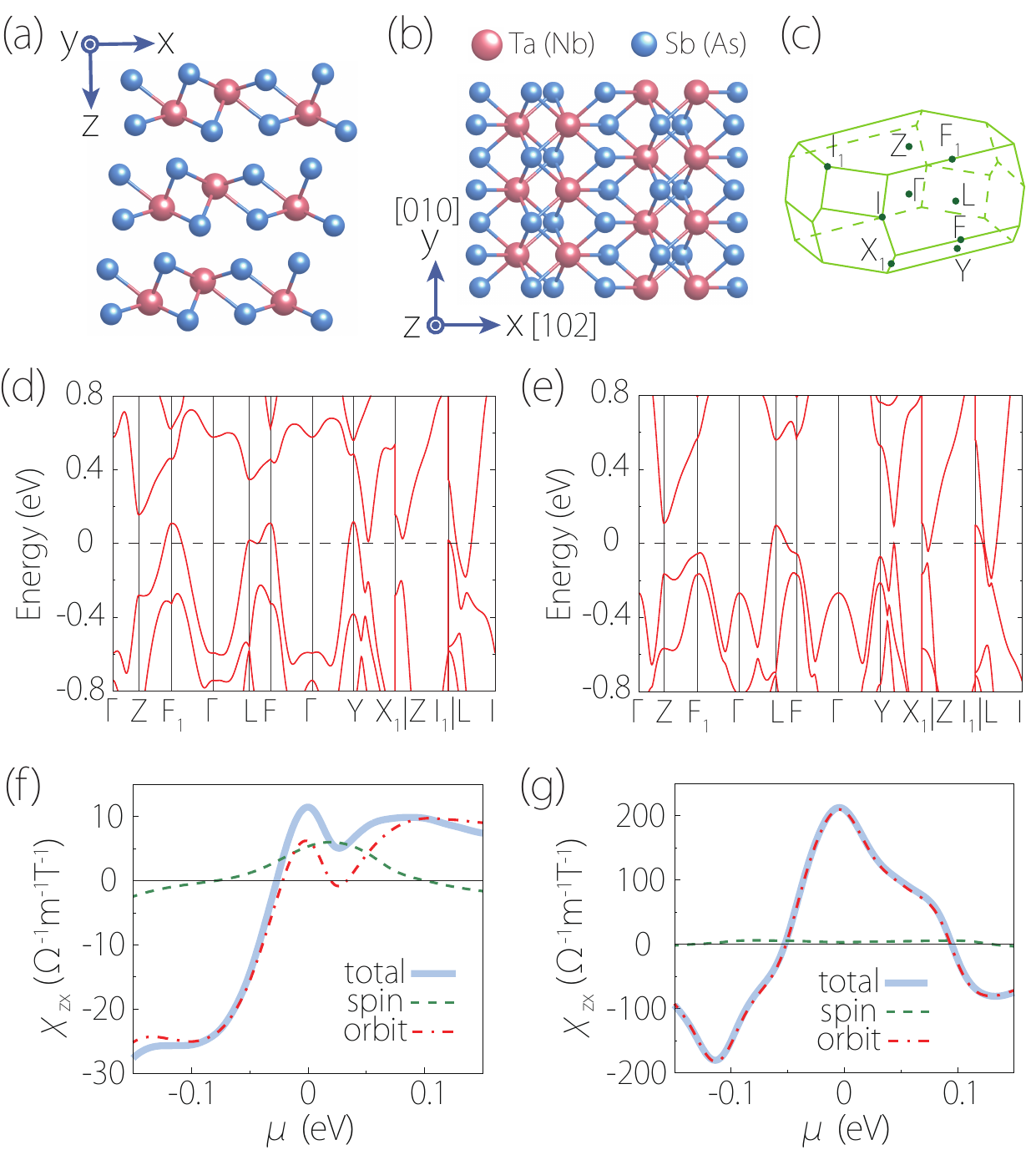} \caption{(a,b) The common crystal
structure of $\mathrm{TaSb_{2}}$ and $\mathrm{NbAs_{2}}$. The cleavage plane
($\bar{2}01$) is taken as the $x$-$y$ plane. (c) shows the Brillouin zone. (d,e) Calculated band
structures (SOC included) for (d) $\mathrm{TaSb_{2}}$ and (e) $\mathrm{NbAs_{2}}$.
(f,g) Calculated IPHE coefficient $\mathcal{X}_{zx}$ for (f)
$\mathrm{TaSb_{2}}$ and (g) $\mathrm{NbAs_{2}}$, as a function of Fermi
energy $\mu$. Here, besides the total result, we also separately plot the spin and orbital contributions. }%
\label{fig1}%
\end{figure}

\textcolor{blue}{\textit{Application to TaSb$_{2}$ and NbAs$_{2}$.}} To
better understand features of IPHE, especially the
relative importance of spin and orbital contributions, we first apply
our theory to two real materials: $\mathrm{TaSb_{2}}$ and $\mathrm{NbAs_{2}}$, belonging to the family of transition metal dipnictides. The two materials are isostructural, but they differ in the strength of SOC: with heavier elements, $\mathrm{TaSb_{2}}$ has a larger SOC strength than $\mathrm{NbAs_{2}}$.
Their common lattice structure is shown
in Fig.~\ref{fig1}(a) and \ref{fig1}(b), which has space group $C2/m$ (No.~12) and
point group $C_{2h}$, with the twofold rotation axis along the $y$
direction here. The two materials have been synthesized by chemical vapor transport method~\cite{Li_2016,Wang_2016}, and their transport and optical properties have been studied in several recent experiments~\cite{Ni2016,Jia2016,Li_2016,shao_2019}.

We perform first-principles calculations on these materials (details are given in Supplemental Material \cite{supp}). The obtained band structures are shown in Fig.~\ref{fig1}(d) and \ref{fig1}(e), which agree with previous studies~\cite{Xu2016}. One can see that
both materials are metallic, and they have a band near-degeneracy region around Fermi level along the $L$-$I$ path. The smallest local vertical gap is around 61.2 (15.6) meV for
$\mathrm{TaSb_{2}}$ ($\mathrm{NbAs_{2}}$). In fact, without SOC, the local gap would close and the two bands would cross~\cite{Xu2016}. The smaller local gap in $\mathrm{NbAs_{2}}$ reflects its weaker SOC than $\mathrm{TaSb_{2}}$.

Recent experiments show the ($\bar{2}01$) plane as the cleavage surface~\cite{Lou_2022,Wadge_2022}, so we take it as the transport plane, which corresponds to the $x$-$y$ plane in Fig.~\ref{fig1}(a). According to Table~\ref{symmetry}, this setup allows only a single $\mathcal{X}_{zx}$ element, and the response can be captured by
\begin{equation}
\mathcal{X}^{\text{H}}=\mathcal{X}_{zx}\cos\varphi. \label{angular}%
\end{equation}
This angular dependence can be directly verified in experiment.

Since our theory is formulated in terms of intrinsic band quantities, it can be easily implemented in first-principles calculations to evaluate the $\mathcal{X}$ tensor.
Figures~\ref{fig1}(f) and \ref{fig1}(g) show the calculated $\mathcal{X}_{zx}$ for $\mathrm{TaSb_{2}}$ and $\mathrm{NbAs_{2}}$, as a function of Fermi energy $\mu$. One can see that the effect is pronounced when
$\mu$ approaches the aforementioned small-gap regions, where the band geometric quantities are enhanced. At the intrinsic Fermi level (i.e., $\mu=0$), we obtain $\mathcal{X}_{zx}=11.45$ $\Omega
^{-1}$m$^{-1}$T$^{-1}$ for $\mathrm{TaSb_{2}}$ and $\mathcal{X}_{zx}=209.92$
$\Omega^{-1}$m$^{-1}$T$^{-1}$ for $\mathrm{NbAs_{2}}$. These values are quite large and are definitely detectable in experiment \cite{Zhou2022}.

\begin{table}[pt]
\caption{{} Calculated IPHE coefficient $\mathcal{X}%
_{zx}$  (in unit of $\Omega^{-1}%
$m$^{-1}$T$^{-1}$) for three concrete materials. For $\mathrm{TaSb_{2}}$
and $\mathrm{NbAs_{2}}$, the transport plane  is set as ($\bar{2}01$). For $\mathrm{SrAs_{3}}$, the transport plane is set as (001). The spin and orbital contributions are also listed separately.}%
\label{material}%
\setlength{\tabcolsep}{2.5mm}{
\begin{tabular}
[b]{ccccc}\hline\hline
\ \ $\mathcal{X}_{zx}$ \ \ & \ \ \ \ $\mathrm{TaSb_{2}}$\ \ \ \  &
\ \ \ \ $\mathrm{NbAs_{2}}$\ \ \ \  & \ \ \ \ \  $\mathrm{SrAs_{3}}$
\\\hline
\ \ spin\ \ \  & 5.32 & 3.00 & \ \ \ \ 4.56 & \\
\ \ orbital\ \ \  & 6.13 & 206.92 & \ \ \ -143.45 & \\
\ \ total\ \ \  & 11.45 & 209.92 & \ \ \ -138.89 & \\\hline\hline
\end{tabular}}
\end{table}

To assess the relative importance of spin and orbital contributions to IPHE, in Figs.~\ref{fig1}(f) and \ref{fig1}(g) we also separately plot
the two contributions. One finds that for $\mathrm{TaSb_{2}}$ with a stronger SOC, the two
contributions are comparable near the intrinsic Fermi level. In contrast, for $\mathrm{NbAs_{2}}$ with a weaker SOC, the orbital
contribution is overwhelmingly dominating over the spin part. The specific values at $\mu=0$
are listed in Table~\ref{material}. These results demonstrate that (1) orbital contribution to IPHE is significant;
(2) beyond previous theories, large IPHE can occur in materials with weak SOC and is dominated by the orbital mechanism.

\begin{figure}[t]
\centering
\includegraphics[width=0.40\textwidth]{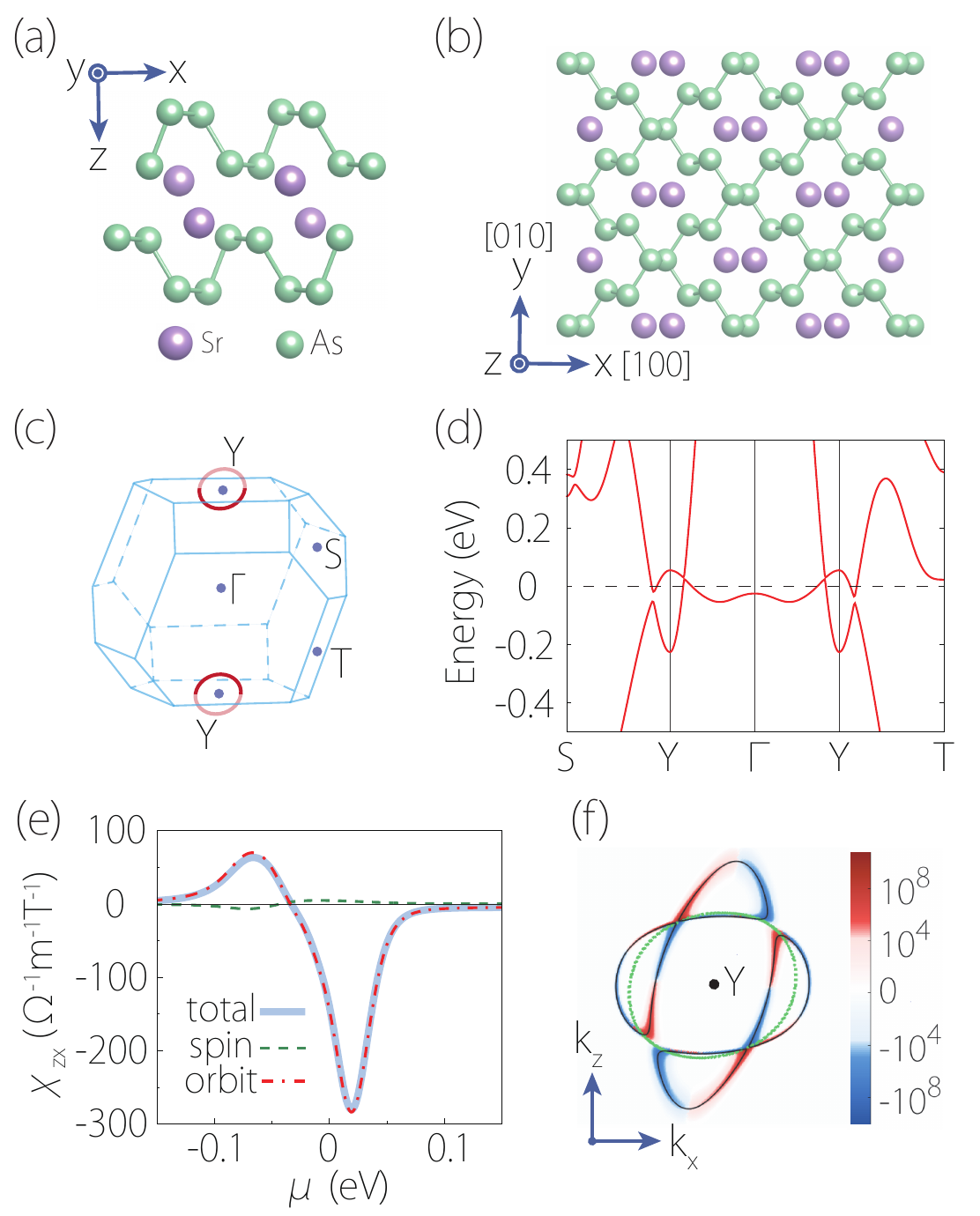} \caption{(a,b) Crystal
structure of $\mathrm{SrAs_{3}}$. Here, the (001) plane is taken as the $x$-$y$ plane. (c) shows its
Brillouin zone. The red line illustrates the nodal loop in the absence of SOC.  (d) Calculated band structure of
$\mathrm{SrAs_{3}}$ (SOC included). (e) Calculated $\mathcal{X}_{zx}$ as a function of $\mu$. (f) $k$-resolved contribution to $\mathcal{X}_{zx}$ (the integrand of Eq.~(\ref{eq-Kai})) on the intrinsic Fermi surface in the $\Gamma$-$Y$-$S$ ($k_{y}=0$) plane. The black lines show the Fermi surface. The green dotted line indicates the original nodal loop in the absence of SOC.
The unit of colormap is $\mathrm{\mu_{B}\mathring{A}^{2}/eV}$.
}%
\label{fig3}%
\end{figure}

\textcolor{blue}{\textit{Application to SrAs$_{3}$.}} The other example we wish to discuss is
the topological semimetal $\mathrm{SrAs_{3}}$. It also has space group $C2/m$,
and its crystal structure is shown in Fig.~\ref{fig3}(a) and \ref{fig3}(b).
Previous works showed that in the absence of SOC, $\mathrm{SrAs_{3}}$ possesses a nodal loop
across the Fermi level in the $\Gamma$-$Y$-$S$ plane [Fig.~\ref{fig3}(c)]~\cite{xu_2017,Wen2018,song_2020}, protected by the mirror symmetry. Inclusion of SOC will gap out the nodal loop, but since the SOC strength in SrAs$_{3}$ is weak, the opened gap is small. From our calculated band structure in Fig.~\ref{fig3}(d), we find the gap values being 32.0 meV and 5.9 meV on $S$-$Y$ and
$Y$-$\Gamma$ paths, respectively, which agree with previous calculations~\cite{song_2020}.
In addition, by scanning around the original nodal loop, we find the smallest gap opened
is $\sim 2.2$ meV. Such small gaps near the Fermi level are expected to generate a large orbital contribution to IPHE.

We take the (001) plane ($x$-$y$ plane in Fig.~\ref{fig3}(a) and \ref{fig3}(b)), which is a cleavage plane of $\mathrm{SrAs_{3}}$~\cite{song_2020}, to be transport plane. Similar to $\mathrm{TaSb_{2}}$ and $\mathrm{NbAs_{2}}$, for IPHE, there is only one nonzero element $\mathcal{X}%
_{zx}$, and the angular dependence follows Eq.~(\ref{angular}). Our calculation result for $\mathcal{X}%
_{zx}$ is plotted in Fig.~\ref{fig3}(e). Again, one observes that in such a material with weak SOC, the effect is dominated by the orbital contribution. Moreover, we find that the orbital contribution is mostly from the AOP dipole \cite{supp}.
At the intrinsic Fermi level, we get $\mathcal{X}%
_{zx}=-138.89$ $\Omega^{-1}$m$^{-1}$T$^{-1}$, which is comparable to the value
in $\mathrm{NbAs_{2}}$.

To correlate this large IPHE with the band topology, in Fig.~\ref{fig3}(f), we plot the integrand of Eq.~(\ref{eq-Kai}) on the
Fermi surface in the $k_{y}=0$ plane, i.e., the plane that contains the original nodal loop. Here, the Fermi surface is marked by the black lines, forming two figure-eight parts, and the green dotted loop indicates the original nodal loop in this plane. One can see that the large contribution is indeed concentrated at regions where the Fermi surface touches the nodal loop.

\textcolor{blue}{\textit{Discussion.}} We have developed a general theory for the IPHE and, importantly, discovered the orbitally induced contribution to the effect. The finding greatly extends the scope of IPHE, which was previously limited only to spin-orbit-coupled systems.
From our calculation results, orbital IPHE can be comparable to spin contribution in systems with strong SOC, and it dominates in systems with weak SOC. Our theory clarifies the band origin of IPHE, especially highlights the role of AOP and ASP, making the effect a new intrinsic property to characterize materials and to probe band geometric quantities.

Orbital IPHE can be directly probed in materials with weak SOC, such as NbAs$_2$ and SrAs$_3$ discussed here. In fact, our analysis~\cite{supp} shows that orbital IPHE  is very likely also dominating the signals reported for ZrTe$_5$~\cite{Ong2018} and $\mathrm{VS_{2}}$-$\mathrm{VS}$~\cite{Zhou2022}, although there are still some uncertainties about these two materials that need to be clarified in experiment, as discussed in Supplemental Material~\cite{supp}.

In this work, we focus on the intrinsic effect, which can be quantitatively evaluated for each material and serves as benchmark for experiment. There should also exist extrinsic contributions arising from disorder scattering. By adopting certain disorder models, they may be evaluated by approaches similar to the anomalous Hall effect \cite{Sinitsyn2007,Nagaosa2010}.
Experimentally, extrinsic effects can be separated by their different scaling with respect to system parameters, such as temperature and disorder strength \cite{Tian2009,Hou2015,Du2019,Xiao2019scaling,Lai2021}.

Recently, there were also studies on nonlinear planar Hall effect, which scales as $\sim E^2B$ \cite{He2019NPHE,Huang2023INPHE,Dantas2023}. Its symmetry property is different from IPHE here. In practice, the responses with different $E$ dependence can be readily
distinguished by applying a low-frequency modulation and using the lock-in technique \cite{Ma2019NLHE,Kang2019,Lai2021}.

Finally, although our theory is developed for nonmagnetic materials, it also applies to
magnetic systems. The main difference there is that regarding intrinsic transport, besides this $B$-linear IPHE, there also exists the conventional $B$-independent intrinsic anomalous Hall effect. In the low-field regime, where the effect of applied $B$ field on magnetic ordering can be treated perturbatively, the two can be readily separated by their different $B$ scaling.

\bibliographystyle{apsrev4-2}
\bibliography{Bilinear_ref}

\end{document}